\documentclass{aa}  

\usepackage{graphicx}
\usepackage{txfonts}
\usepackage{hyperref}

\hypersetup{
    final=true,
    pageanchor=true,
    colorlinks=true,
    breaklinks=true,
    linkcolor=blue,
    citecolor=blue,
    urlcolor=blue,
    pdfpagemode=UseNone}

\usepackage[switch, modulo]{lineno}

\usepackage{natbib}
\bibpunct{(}{)}{;}{a}{}{,}
\usepackage{multirow}
\usepackage{nicefrac}
\usepackage[usenames,dvipsnames]{color}

\definecolor{blue}{rgb}{0.00, 0.00, 1.00}
\definecolor{red}{rgb}{0.86, 0.08, 0.24}
\definecolor{green}{rgb}{0.00, 0.39, 0.00}

\newcommand\kms{km\,s$^{-1}$}

\begin{document} 

  \title{Chromospheric velocities in an M3.2 flare using \ion{He}{i} 1083.0 nm and \ion{Ca}{ii} 854.2 nm}

   \author{C. Kuckein\inst{1,2}
          \and
          M. Collados\inst{1,2}
          \and 
          A. Asensio Ramos\inst{1,2}
          \and
          C. J. D\'iaz Baso\inst{3,4}
          \and
          T. Felipe\inst{1,2}
           \and
          C. Quintero Noda\inst{1,2}
                 \and
          L. Kleint\inst{5}
                 \and
          L. Fletcher\inst{6,3,4}
                 \and
          S. Matthews\inst{7}
          }

   \institute{
    Instituto de Astrof\'isica de Canarias (IAC), V\'ia L\'actea s/n, E-38205 La Laguna, Tenerife, Spain  \\
    \email{ckuckein@iac.es}        
    \and
    Departamento de Astrof\'\i sica, Universidad de La Laguna, E-38206 La Laguna, Tenerife, Spain
    \and
    Institute of Theoretical Astrophysics,
    University of Oslo, %
    P.O. Box 1029 Blindern, N-0315 Oslo, Norway
    \and
    Rosseland Centre for Solar Physics,
    University of Oslo, %
    P.O. Box 1029 Blindern, N-0315 Oslo, Norway
    \and
    Astronomical Institute of the University of Bern, Sidlerstrasse 5, 3012 Bern, Switzerland
    \and
    School of Physics and Astronomy, University of Glasgow, Glasgow G12 8QQ, UK
    \and
    UCL Mullard Space Science Laboratory, Holmbury St Mary, Dorking RH5 6NT, UK
          }

   \date{Version: \today }

 
  \abstract
   {} 
   {We aim to study the chromospheric line-of-sight (LOS) velocities during the GOES M3.2 flare (SOL2013-05-17T08:43) using simultaneous high-resolution ground-based spectroscopic data of the \ion{He}{i} 10830\,\AA\ triplet and \ion{Ca}{ii} 8542\,\AA\ line. A filament was present in the flaring area.}
   {The observational data were acquired with the Vacuum Tower Telescope (VTT, Tenerife, Spain) and covered the pre-flare, flare, and post-flare phases. 
   Spectroscopic inversion techniques (HAZEL and STiC) were applied individually to \ion{He}{i} and \ion{Ca}{ii} lines to recover the atmospheric parameters of the emitting plasma.
   Different inversion configurations were tested for \ion{Ca}{ii} and two families of solutions were found to explain the red-asymmetry of the profiles: a redshifted emission feature or a blueshifted absorption feature. These solutions could explain two different flare scenarios (condensation vs. evaporation). The ambiguity was solved by comparing these results to the \ion{He}{i} inferred velocities. }
   {At the front of the flare ribbon, we observed a thin, short-lived blueshifted layer. This is seen in both spectral regions but is much more pronounced in \ion{He}{i}, with velocities of up to $-10$\,\kms. In addition, at the front we found the coexistence of multiple \ion{He}{i} profiles within one pixel. The central part of the ribbon is dominated by \ion{He}{i} and \ion{Ca}{ii} redshifted emission profiles. A flare-loop system, visible only in \ion{He}{i} absorption and not in \ion{Ca}{ii},  becomes visible in the post-flare phase and shows strong downflows at the footpoints of up to 39\,\kms. In the flare, the \ion{Ca}{ii} line represents lower heights compared to the quiet Sun, with peak sensitivity shifting from $\log \tau \simeq -5.2$ to $\log \tau \simeq -3.5$. The loop system's downflows persist for over an hour in the post-flare phase.}
   {The inferred LOS velocities support a cool-upflow scenario at the leading edge of the flare, with rapid transition from blueshifts to redshifts likely to occur within seconds to tens of seconds. 
   Although the flare had a significant impact on the surrounding atmosphere, the solar filament in the region remained stable throughout all flare phases. 
   The inclusion of the \ion{He}{i} triplet in the analysis helped resolve the ambiguity between two possible solutions for the plasma velocities detected in the \ion{Ca}{ii} line. This highlights the importance of combining multiple chromospheric spectral lines to achieve a more comprehensive understanding of flare dynamics.
   }

   \keywords{Sun: flares --
             Sun: chromosphere --
             Sun: activity -- 
             Methods: data analysis --
             Techniques: spectroscopic }

   \authorrunning{Kuckein et al.}
   \titlerunning{Chromospheric velocities in an M3.2 flare using \ion{He}{i} 1083.0 nm and \ion{Ca}{ii} 854.2 nm}      
             
   \maketitle

\section{Introduction}
Solar flares are among the most energetic phenomena in our solar system, characterized by a sudden and intense brightening across multiple wavelength ranges in the solar spectrum. These events are triggered in the corona by the reconfiguration of the magnetic field, resulting in the rapid release of enormous amounts of stored magnetic energy \citep{priest02,fletcher11}. During flares, this energy is converted into thermal energy, radiation across the electromagnetic spectrum, and kinetic energy manifested as plasma flows and particle acceleration. Understanding the dynamics of these plasma flows is crucial to understand the mechanisms of energy transport and dissipation during flares.

The chromosphere plays an important role during flares \citep{hudson07,milligan15}. A significant part of the flare energy goes into continuum radiation, but the exact distribution of the energy conversion is not well known \citep[][]{milligan14,kleint16}. This chromosphere undergoes dramatic changes in its thermodynamic properties and flow patterns during flares. Chromospheric evaporation and condensation -- the upward and downward flows of plasma in response to the flare -- have been studied using spectroscopic observations \citep{zarro88, keys11, yadav21}. For instance, the H$\alpha$ line exhibits red and blueshifts of 17\,\kms\ in a C-class flare \citep{keys11}.

Among the most popular chromospheric lines for flare-analysis are the H$\alpha$ and \ion{Ca}{ii} 8542\,\AA\, lines, as well as the \ion{He}{i} 10830\,\AA\ triplet. In this work we concentrate on the latter two. 
The \ion{Ca}{ii} 8542\,\AA\ spectral line is of particular interest, as it has been often used to study flares \citep{harvey12, kleint12, kleint17, kuridze17, kuridze18,Yadav2022,vissers21,ferrente23,ferrente24}. It is formed under non-local thermodynamical equilibrium (NLTE) conditions and originates from the transition between the upper $^2$P$_{1/2}$ and lower $^2$D$_{3/2}$ levels of the ionized \ion{Ca}{ii} atom. 
The line is sensitive to plasma conditions from the lower photosphere to the middle chromosphere \citep[e.g.,][]{quinteronoda16,kuckein17}.
However, simulations have shown that during flares the typical formation height of the core of 8542\,\AA\ shifts lower in height by a few hundreds of kilometers \citep{dacosta15,kerr16}. The \ion{He}{i} 10830\,\AA\ triplet, which forms in the upper chromosphere \citep{avrett94}, provides complementary information about higher atmospheric layers and often shows emission profiles followed by strong absorption with significant Doppler shifts during flares \citep{penn95,kuckein15a, kuckein15b, judge15, anan18}. Recent space-based high-resolution observations have revealed the spatial and temporal complexity of these flows, including the discovery of localized regions at the leading edges of flare ribbons exhibiting oppositely directed flows compared to the main flare areas \citep{xu16,tei18,panos18,polito23}.

\begin{figure*}[!t]
 \centering
\includegraphics[width=0.95\hsize]{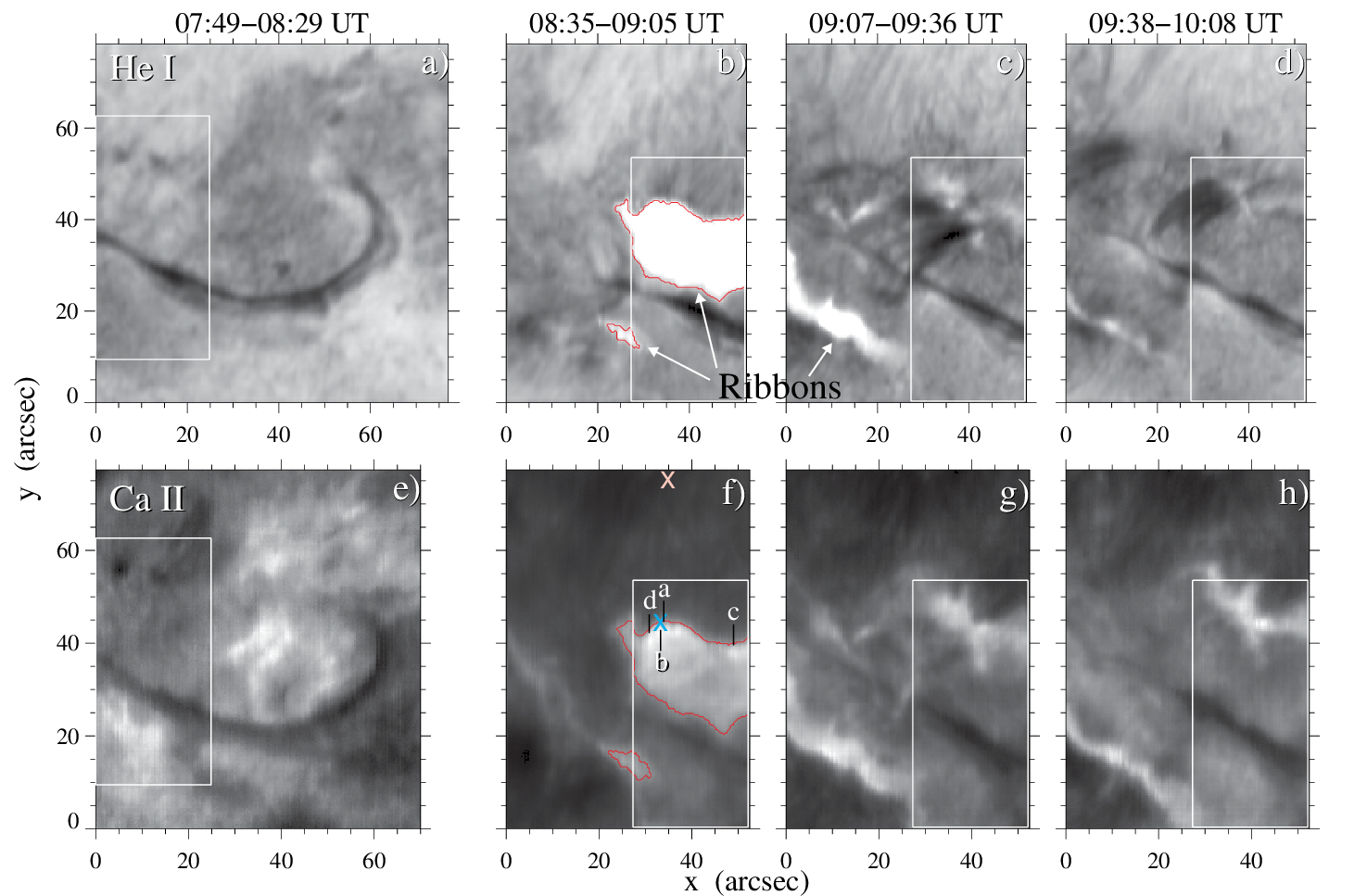}
\caption{Slit-reconstructed images of the four raster scans centered at the \ion{He}{i} line core of the red component (upper row) and the \ion{Ca}{ii} line core (lower row). From left to right: pre-flare (panel `a)' and `e)'), flare (panel `b)' and `f)'), and post-flare maps (panels `c)', `d)', `g)', and `h)'). The white box shows the common FOV among all raster scans. The letters a--d in panel f) mark the position of the pixels shown in Fig. \ref{Fig:Caprofiles}. The light blue and pink crosses in panel f) mark the positions of the response functions shown in Fig. \ref{Fig:RF}. The red contours mark the borders of the flare ribbons in panels b) and f), using an intensity threshold of $I/I_\mathrm{c} = 0.87$ in the line core. } 
  \label{Fig:VTTallmaps}
\end{figure*}

\begin{figure}[!t]
 \centering
 \includegraphics[width=\hsize]{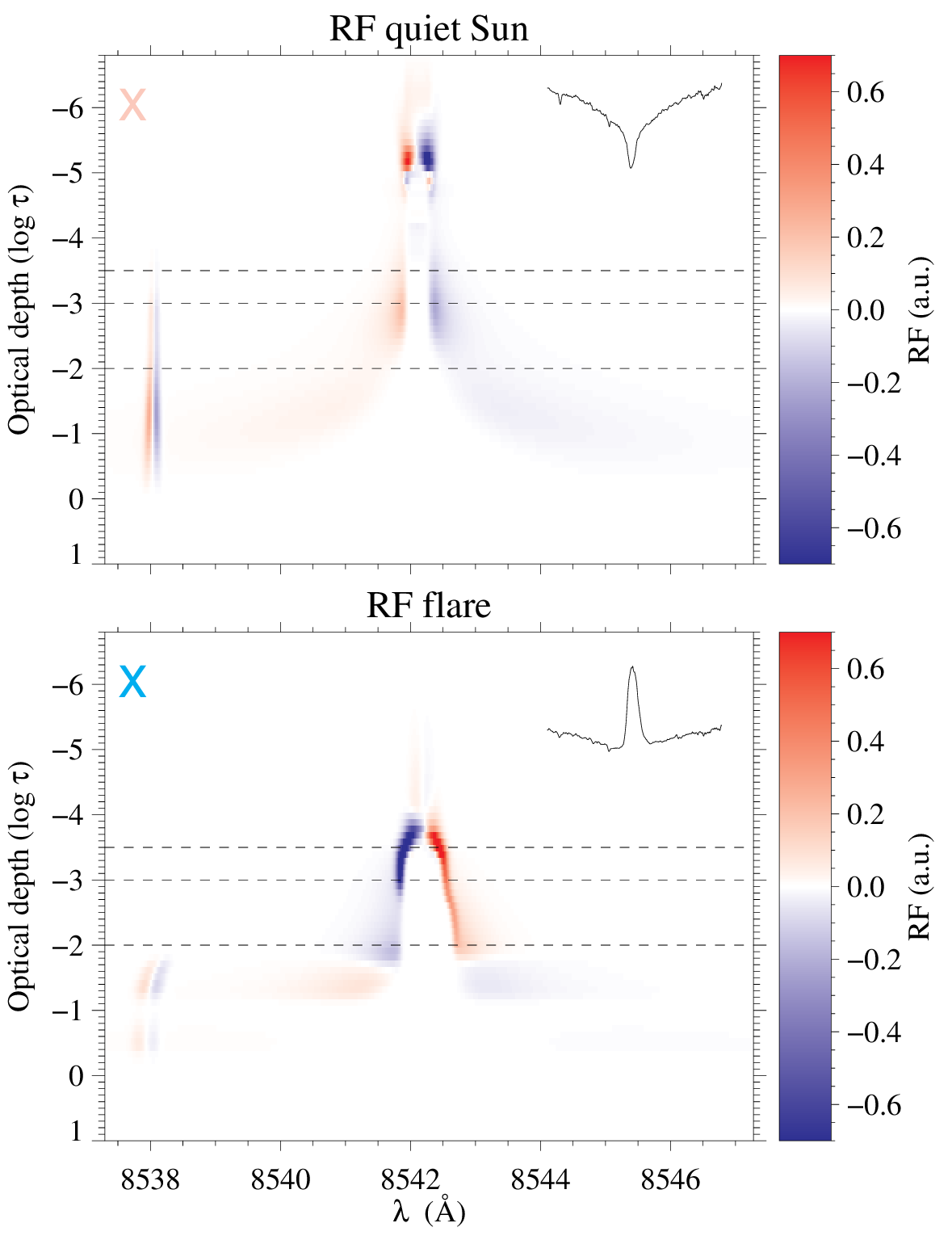}
\caption{Response functions of the \ion{Ca}{ii} 8542\,\AA\ spectral line to changes in the LOS velocity. The lower panel corresponds to a pixel within the bright flare, at the ribbon front at coordinates $x,y = 33\farcs9,44\farcs6$ (see light-blue cross in Fig. \ref{Fig:VTTallmaps}f), whereas the upper panel shows the RF in a quiet-Sun pixel (pink cross in Fig. \ref{Fig:VTTallmaps}f at $x,y = 35\farcs0,75\farcs2$). The RFs are normalized to their maximum absolute value and saturated between $\pm0.7$ to enhance the weaker areas. The corresponding observed intensity profile is shown in the upper right corner of each panel and extends over the same spectral range as the RF. The horizontal dashed lines mark the optical depths at $\log \tau = -2.0$, $-3.0$, and $-3.5$.  } 
  \label{Fig:RF}
\end{figure}

Spectral line inversion tools, such as NICOLE \citep{socas15}, STockholm inversion Code \citep[STiC][]{stic}, DeSIRe \citep{desire}, and ``HAnle and ZEeman Light''  \citep[HAZEL][]{hazel}, are capable of deriving physical parameters from the \ion{Ca}{ii} 8542\,\AA\ line or \ion{He}{i} 10830\,\AA\ triplet, providing valuable information about the atmosphere where the line originated. It is important to note that Stokes profiles in flares are often highly complex. 
The stratification inferred from single-line studies is often poorly constrained, as different solutions can reproduce the same spectra. Therefore, multi-line studies are necessary for an accurate interpretation \citep[e.g.,][]{DiazBaso_2021}, and we aim to demonstrate this in our work.

The relationship between flare energy deposition and the resulting plasma flows is still not fully understood. Numerical models suggest that the magnitude and direction of flows depend on the depth and rate of energy deposition \citep{fisher85,reep15}, but comprehensive observational constraints combining multiple spectral lines formed at different heights remain relatively scarce. Additionally, the temporal evolution of flows from the impulsive phase to the gradual phase reveals important information about the flare energy transport mechanisms. By combining observations from different chromospheric lines, we can build a more complete picture of the height-dependent velocity structure during flares.

In this paper, we focus on the analysis of high-resolution ground-based spectroscopic data of \ion{He}{i} 10830\,\AA\ and  \ion{Ca}{ii} 8542\,\AA\ during different phases of an M-class flare. We describe the intensity profiles and infer the line-of-sight (LOS) velocities during the flare.

\section{Observations}
We followed the evolution of active region NOAA 11748 between 07:49\,UT and 10:08\,UT of 2013 May 17 with the Vacuum Tower Telescope \citep[VTT,][]{vtt} located on Tenerife, Spain. The main target was an AR filament lying on top of the polarity inversion line (PIL) at solar disk coordinates ($x$,$y$) $\sim$ ($-$540\arcsec, 200\arcsec) and cosine of the heliocentric angle $\mu = \cos \theta\sim 0.8$. During the observations, an M3.2 flare started at 08:43\,UT (SOL2013-05-17T08:43:00) and peaked at 08:57\,UT. The two-ribbon flare quickly evolved on both sides of the filament and expanded across the whole active region (see Fig. \ref{Fig:VTTallmaps}).

The setup included the Echelle spectrograph from the VTT to acquire high-resolution intensity spectra centered at the infrared \ion{Ca}{ii} 8542~\AA\ line with a PCO.4000 CCD camera. The width of the slit was 100~$\mu$m and the scanning step 0\farcs35. The slit was oriented along the terrestrial North--South direction. The pixel size along the slit was 0\farcs17. The exposure time for the \ion{Ca}{ii} spectra was 1\,s. The Kiepenheuer-Institute Adaptive Optics System \citep[KAOS;][]{Berkefeld10} was locked on high-contrast structures, such as pores and small penumbrae, and was crucial to improve the observations. 

Spectropolarimetric data acquired simultaneously with the Tenerife Infrared Polarimeter \citep[TIP-II;][]{tip2} in the \ion{Si}{i} 10827~\AA\ and \ion{He}{i} 10830~\AA\ spectral region was analyzed by \citet[][]{kuckein15a, kuckein15b}. However, the chromospheric LOS velocities of \ion{He}{i} were not studied. In this work, we use the data to study the flows during the present flare, combining different chromospheric lines. The details of the \ion{He}{i} 10830~\AA\ observations are described in \citet[][]{kuckein15a}. The key points are mentioned here: the exposure time per slit position was 10\,s, and the scanning step and pixel size were the same as for the abovementioned \ion{Ca}{ii} 8542~\AA\ observations. The observed spectral range spanned between 10824 and 10835\,\AA, with a spectral sampling of about 11.1\,m\AA\,px$^{-1}$.

Four raster scans, covering almost continuously the time between 07:49 and 10:08\,UT, were carried out.  These maps cover the pre-flare (map 1), impulsive (map 2), and relaxation phases (maps 3 and 4) of the flare. The first map encompassed 220 and 200 steps, for the \ion{He}{i} 10830~\AA\ and \ion{Ca}{ii} 8542~\AA\ observations, respectively.  Maps 2, 3, and 4 covered 150 steps with the slit in about 30\,min each. A two-pixel binning along the slit was carried out on all data to produce squared pixels. The coverage and common FOV of the maps are depicted as white rectangles in Fig.~\ref{Fig:VTTallmaps}.

\section{Data reduction and analysis}
\subsection{\ion{He}{i} 10830\,\AA\ calibration}
The data was corrected for dark-current and flat-field variations across the detector, as well as the standard polarimetric calibration \citep[][]{collados99,collados03}. The normalized Stokes profiles were computed by dividing them by the spatially-averaged continuum in a quiet-Sun area. The spectral range included two telluric lines, which were used for an absolute wavelength calibration as described by \citet{kuckein12b} and \citet{martinez97}. A three-pixel binning was performed in the spectral dimension, to enhance the signal-to-noise ratio. The spectral sampling was therefore $\sim$33.3\,m\AA\,px$^{-1}$. 

\subsection{\ion{Ca}{ii} 8542\,\AA\ calibration}
Standard dark-current and flat-field corrections were carried out for the \ion{Ca}{ii} spectra. An average spectrum was computed to compensate the prefilter transmission. 
A three-pixel binning in the spectral dimension was performed to enhance the signal-to-noise ratio. 
The final wavelength grid comprised 400 positions spanning 10\,\AA\ with an equidistant spectral step of 25\,m\AA\,px$^{-1}$.
In addition, the spectra were slightly smoothed by convolving them with a normalized Gaussian with a FWHM of about 50~m\AA. As a final step, we performed an intensity calibration of each map  by scaling the intensity in a quiet-Sun area to the spectra from the Fourier Transform Spectra \citep[FTS,][]{neckel84} spectrometer and taking into account the observed heliocentric angle following \cite{Allen}, as described in  \cite{DiazBaso_2019}. After this, a final inspection showed a residual behavior where the blue wing of the \ion{Ca}{ii} 8542\AA\ line exhibited systematically a lower intensity compared to the corresponding continuum at the red side. To address this issue, a solution was implemented by fitting a parabolic function to the ratio of the atlas and the quiet-Sun profile of each scan.
In the absence of clearly defined telluric lines, we carried out the wavelength calibration using two lines in the outer blue wing of the \ion{Ca}{ii} line: (1) the \ion{Si}{i} line at 8536.163~\AA\ and (2) the \ion{Fe}{i} line at 8538.021~\AA\ \citep[wavelengths taken from][]{moore66}.

\subsection{\ion{He}{i} 10830\,\AA\ inversions}\label{Sect:inversionsHe}
The four Stokes parameters were inverted with the inversion code  HAZEL\footnote{\url{https://github.com/aasensio/hazel2}} to retrieve a model atmosphere. The inversion strategy involved a single chromospheric slab and four cycles, that is, four repetitions of the inversion process per pixel, each using the results of the previous cycle. The first cycle only took into account the intensity profile, to find a first solution for the optical depth, Doppler velocity, Doppler width, damping parameter, and a parameter $\beta$ (see below), which will be improved in the following cycles. The remaining three cycles included Stokes $Q$, $U$, and $V$ to find a more accurate solution for the above-mentioned parameters taking into account the magnetic field. The $\beta$ parameter plays an important role when fitting emission profiles on the solar disk. This parameter was introduced ad-hoc into HAZEL and multiplies the source function $S$ to produce emission of the \ion{He}{i} 10830\,\AA\ triplet. HAZEL fits the whole spectral range, including the neighboring \ion{Si}{i} 10827\,\AA\ and telluric line, in order to account for strong Doppler shifts of the \ion{He}{i} triplet. The results of the magnetic field are beyond the scope of this work and will be postponed for a future publication.

\subsection{\ion{Ca}{ii} 8542\,\AA\ inversions}\label{Sect:inversionsCa}
We estimated the model atmosphere for the \ion{Ca}{ii} observations through NLTE inversions. The inversion of the \ion{Ca}{ii} 8542\,\AA\ line was performed using the parallel NLTE code STiC\footnote{\url{https://github.com/jaimedelacruz/stic}}, which is built on top of an optimized version of the RH code \citep{rh01}.  Our model was constructed by assuming independent 1D plane-parallel atmospheres along each LOS, an approach commonly referred to as 1.5D modeling.

STiC iteratively adjusts the physical parameters of a model atmosphere, such as the temperature, LOS velocity, magnetic field vector, and microturbulence to find a synthetic spectrum that reproduces the observed profile. The density and gas pressure stratifications are computed by assuming hydrostatic equilibrium (HE). 
Although HE is likely not valid in flares, it remains the best available approximation. Deviations from HE can introduce uncertainties in the optical depth scale, potentially affecting the absolute location of certain atmospheric features. However, we consider the relative variations between atmospheric layers and the overall trends to be reliable, and thus the main conclusions regarding mass flow patterns remain robust. The physical parameters of the model atmosphere are given as functions of the optical depth scale at 5000\,\AA, hereafter $\log \tau$. The physical parameters are modified at specific node locations, followed by an interpolation to all other depth points. The merit function that accounts for the likelihood between the synthetic and observed spectra includes an additional regularization term that ensures a smooth behavior in the atmosphere even when a physical parameter has many degrees of freedom (that is, nodes in the stratification).

We treated the \ion{Ca}{ii} atom in NLTE with the \ion{Ca}{ii}~8542\,\AA\ line in complete frequency redistribution (CRD). Although STiC can invert the adjacent photospheric lines simultaneously, we have not included them. 
The reason is that our initial attempts to include the photospheric lines in the STiC inversions produced unsatisfactory results. In particular, it was difficult to achieve a consistent fit for both the broad \ion{Ca}{ii} 8542\,\AA\ line and the much narrower photospheric lines. This mismatch may be attributed to inaccuracies in atomic parameters, differences in line widths requiring careful weighting in the $\chi^2$ minimization, or noise in the weaker photospheric lines. Given our focus on the lower and middle chromosphere, we restricted the inversions to the \ion{Ca}{ii} line. 
This implies that our sensitivity to the lower photosphere is diminished, so results around and above ($\log\tau=-1$) should be interpreted with caution. Particularly during the flare peak, there are significant gradients in the velocity stratification that might lead some nodes to reflect small potentially spurious negative velocities at the lower levels where sensitivity is reduced.
We initialize the model atmosphere from the FAL-C model \citep{Fontenla1993} by discretizing it in 81 equidistant depth points from $\log \tau=-6.8$ to $\log \tau =+1.2$. Since the observations were acquired without polarimetry, we set the magnetic field vector to zero. To take into account the spectral degradation of our observations, we have used a Line Spread Function (LSF) with a Gaussian shape and a FWHM of 60~m\AA. This value was estimated comparing the FTS spectrum and our final average quiet-Sun profile. During the inversion we weighted all spectral points equally.

Finally, to investigate the robustness of the solution of the inversion problem, we opted to run the inversions using two different configurations: a) in multiple cycles of increasing amount of nodes and b) in multiple cycles starting with a large number of nodes. The only difference is the number of nodes of the LOS velocity. This allowed us to find two different families of solutions that will be discussed in the following section. The inversion strategies are summarized in Table~\ref{Tab:nodes}. We note that the high number of nodes required in these inversions, compared to other contemporary studies \citep[e.g.,][]{kuridze17},
results from the detailed spectral information provided by our slit spectrograph and the dynamic nature of the flare profiles. Nevertheless, the strong regularization applied to the inversion nodes significantly reduces the effective degrees of freedom in the optimization, ensuring a smooth and consistent solution.
In summary, since the observations were acquired without polarimetry, we only have as free parameters the temperature ($T$), the line-of-sight velocity (v$_{\rm LOS}$) and the microturbulent velocity (v$_{\rm turb}$). The same configurations were used for the four maps. 

\begin{table}[!h]
\centering
\small
\begin{tabular}{c|c|c|c}
 & Cycle 1 & Cycle 2 & \\[0.5ex]
\hline\hline\rule[2mm]{0mm}{2.5mm}
Config.~A & $T$=5, v$_{\rm LOS}$=2, v$_{\rm turb}$=1 & $T$=17, v$_{\rm LOS}$=9, v$_{\rm turb}$=2 &   \\[1ex]
Config.~B & $T$=5, v$_{\rm LOS}$=9, v$_{\rm turb}$=1 & $T$=17, v$_{\rm LOS}$=9, v$_{\rm turb}$=2 &   \\[1.5ex]
\end{tabular}
\caption{Number of nodes in each physical parameter for each configuration of the \ion{Ca}{ii} 8542\,\AA\ inversions. }
\label{Tab:nodes}
\end{table}

\begin{figure*}[!t]
 \sidecaption
\includegraphics[width=0.6\hsize]{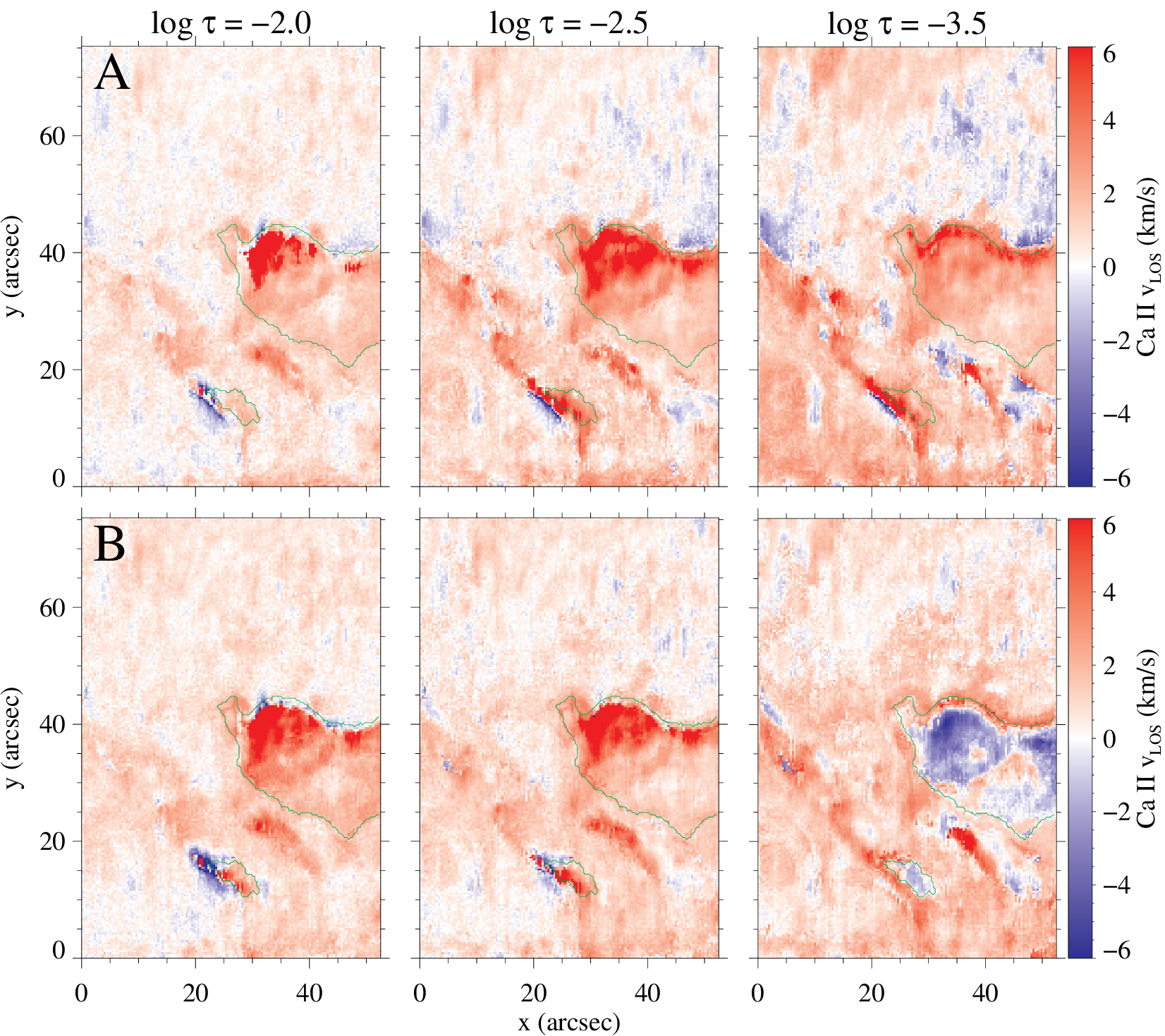}
  \caption{Comparison of the \ion{Ca}{ii} 8542\,\AA\ LOS velocities at different optical depths, using two different inversion strategies with STiC (see Table \ref{Tab:nodes}), during the M-class flare (map 2 or panel f) in Fig. \ref{Fig:VTTallmaps}). Top (A): restrictions applied to the inversion code.  Bottom (B): solution leaving the code with more freedom. The velocities are clipped between $\pm 6$\,\kms. The faint green contours mark the borders of the ribbons as seen in Fig. \ref{Fig:VTTallmaps}f), using an intensity threshold at $I/I_\mathrm{c} = 0.87$.} 
  \label{Fig:caiivelmaps}
\end{figure*}

\section{Results}
In this section, we present the results of the plasma flows in the chromosphere during the M-class flare, as inferred from the Doppler velocities from the \ion{Ca}{ii} 8542\,\AA\ line and \ion{He}{i} 10830\,\AA\ triplet.

\subsection{\ion{Ca}{ii} 8542\,\AA}
The \ion{Ca}{ii} 8542\,\AA\ spectral line yields information from the low to mid chromosphere in the quiet Sun and quiescent active regions. However, it is known that the height of the chromosphere is different/altered during flares. To address this, the response functions (RFs), that is, the sensitivity of the \ion{Ca}{ii} 8542\,\AA\ line to changes in the LOS velocity with height in the atmosphere (given by the optical depth), were computed with STiC. The RFs for a pixel in the quiet Sun and a pixel in the flare ribbon, marked in Fig.~\ref{Fig:VTTallmaps}f, are depicted in Fig.~\ref{Fig:RF}. 
To obtain these, we inverted two pixels and calculated the RFs from the resulting models. These cases are representative, as similar results are found for other profiles with a similar shape.
While the \ion{Ca}{ii} 8542\,\AA\ line is sensitive up to approximately $\log\tau = -5.8$ in the quiet Sun, with peak sensitivity around $\log\tau = -5.2$, its upper optical depth with significant sensitivity extends only to about $\log\tau = -3.8$ during the flare, after which it rapidly decreases. This important change in optical depth must be considered when analyzing the LOS velocity maps.

\begin{figure*}[!t]
\centering
\includegraphics[width=0.95\hsize]{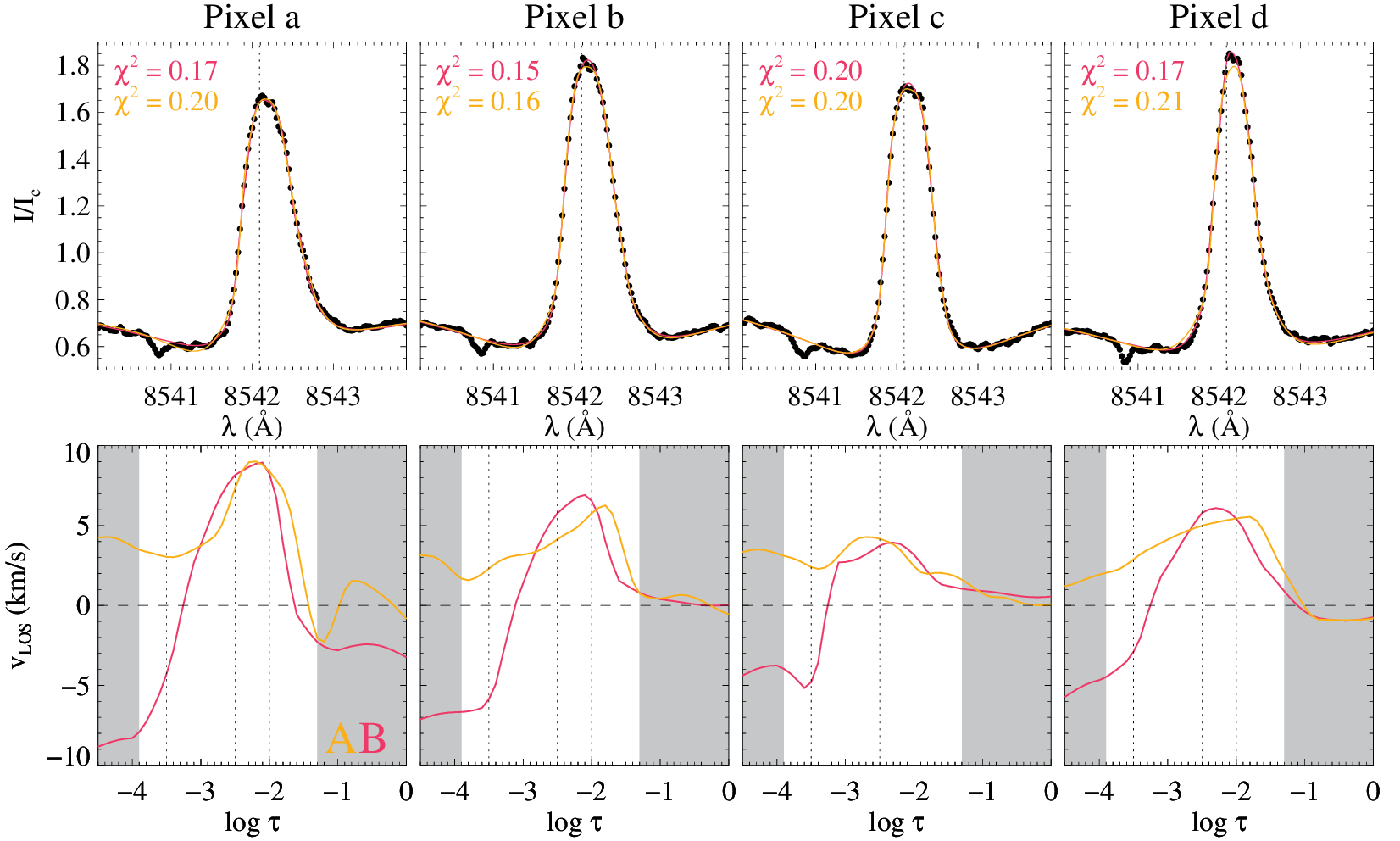}
\caption{Selected \ion{Ca}{ii} 8542\,\AA\ profiles from the flare-front ribbon and inferred LOS velocity stratification. Top: The black dots show the observed \ion{Ca}{ii} intensity profiles, the pink and orange lines depict the fit from the inversions using two different inversion configurations (Table \ref{Tab:nodes}). The $\chi^2$-test shows similar values for both fits. The vertical dotted line marks the center of the \ion{Ca}{ii} line at rest. Bottom: inferred LOS velocity stratification with height on an optical depth scale. The orange (pink) line corresponds to configuration A (B) setup of the inversion (see Table \ref{Tab:nodes}). The vertical dotted lines mark the three optical depths shown in Fig. \ref{Fig:caiivelmaps}. The position of the pixels within the FOV is represented in panel f) of Fig. \ref{Fig:VTTallmaps}. The velocities shown inside the shaded area are not trustworthy, as the RF is very low in that area. 
} 
  \label{Fig:Caprofiles}
\end{figure*}

Given that the RF during the flare peaks at approximately $\log\tau = -3.5$, we have chosen to present the LOS velocities inferred with STiC at $\log\tau = -2.0$, $-2.5$, and $-3.5$ in Fig.~\ref{Fig:caiivelmaps}. Outside this range, the sensitivity of the \ion{Ca}{ii} line to Doppler velocities quickly drops. After several tests, two different inversion strategies (see Table \ref{Tab:nodes}) were used to retrieve the Doppler velocities. Focusing on the flaring area of Fig.~\ref{Fig:caiivelmaps}, the velocities in the upper row (configuration A) change smoothly from the lower atmospheric layers ($\log\tau = -2.0$) to the higher layers ($\log\tau = -3.5$). In contrast, in the lower panels (configuration B), there is an abrupt change in the sign of the velocities between $\log\tau = -2.5$ and $-3.5$. To understand this difference, a few individual example profiles (pixels a--d in Fig. \ref{Fig:VTTallmaps}f), and their fit from the inversions, are shown in Fig.~\ref{Fig:Caprofiles}. The pixels belong to the front of the flare ribbon. The observed profiles appear as black dots, whereas the best fit from the inversions is overplotted as solid orange and pink lines, for the inversions with configuration A and B, respectively. Critically, both models reproduce the observations with comparable precision, indicating the degeneracy of the inversion process. Specifically, while both models show similar redshifts at lower and mid-layers ($\log\tau = -2$ and -2.5) where a strong red-wing asymmetry is observed around 8542.6\AA\, their solutions differ significantly in the higher layers (above $\log\tau = -3.0$). In these higher layers, where the line core is formed, the pink model systematically shows blueshifts, while the orange model exhibits redshifts. The reason why the inversion code converges to two distinct solutions can be attributed to the flexibility inherent in configuration B, which has more freedom during the first inversion cycle compared to configuration A. These results show that the red asymmetry near the line core can be explained by two possible scenarios: a redshifted emission (configuration A) or a blueshifted absorption (configuration B). To determine which scenario is more physically plausible and to understand the velocity structure in these upper layers, we have analyzed the \ion{He}{i} LOS velocities, and the results are presented below.

\begin{figure}[!ht]
 \centering
 \includegraphics[width=0.9\hsize]{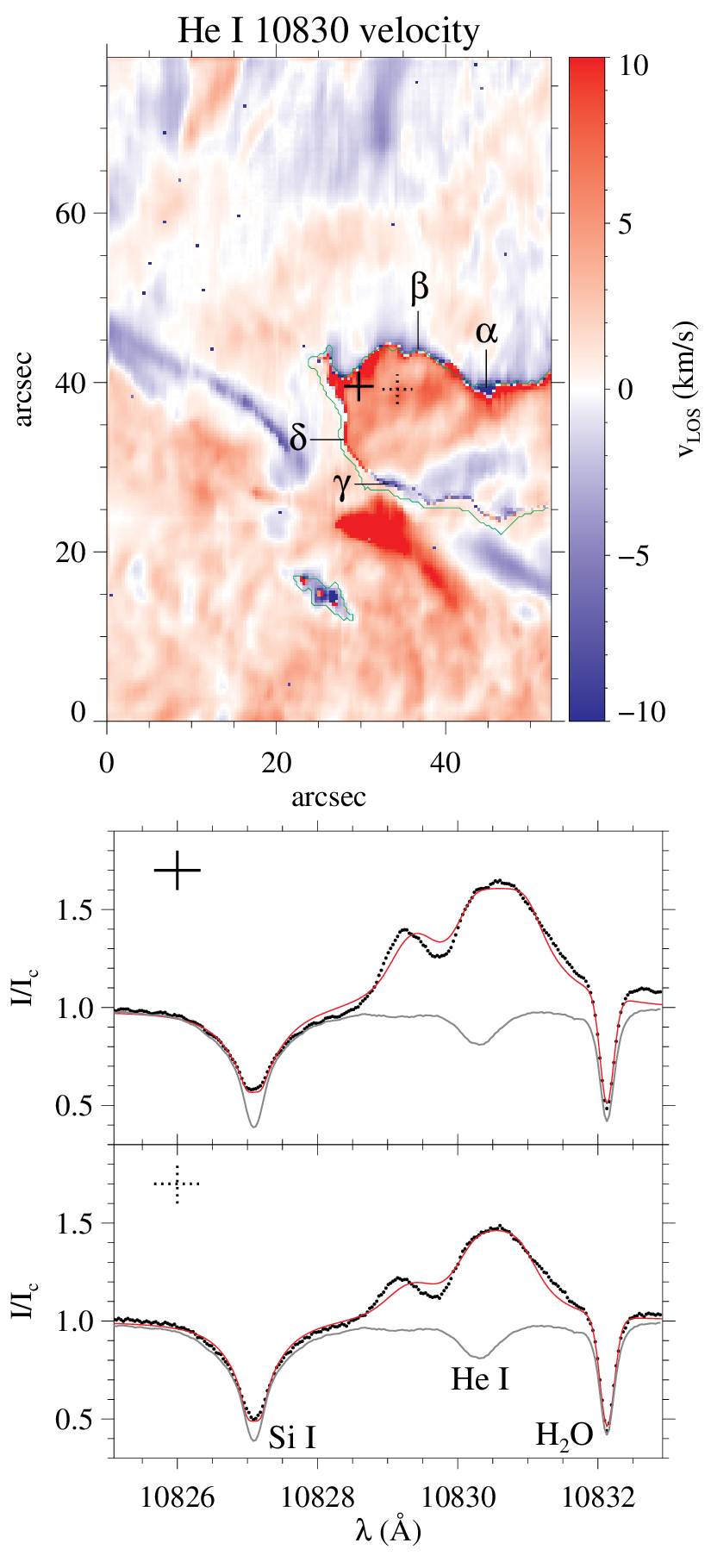}
\caption{Top: slit-reconstructed LOS velocity map of the flare inferred from the \ion{He}{i} 10830\,\AA\ inversions using HAZEL. The map is clipped between $\pm 10$\,\kms. Red (blue) colors indicate flows directed away (toward) the observer. The faint green contour marks the borders of the ribbons as seen in Fig. \ref{Fig:VTTallmaps}b), using an intensity threshold at $I/I_\mathrm{c} = 0.87$. Bottom: Two example intensity profiles observed within the flaring area. The black dots are the observations, whereas the red solid line depicts the fit from the HAZEL inversion. The two profiles are marked inside the velocity map with a solid and dotted cross. The solid- and dotted-cross profiles correspond to an inferred velocity of 8.0 and 6.9\,\kms, respectively, retrieved from the \ion{He}{i} fit. The solid gray line shows the average quiet-Sun profile. Letters $\alpha - \gamma$ show the positions of the leading-edge profiles exhibited in Fig. \ref{Fig:profsbluefront}. } 
  \label{Fig:Hevlos}
\end{figure}

\subsection{\ion{He}{i} 10830\,\AA}
The LOS velocities inferred from the \ion{He}{i} triplet during the flare (map 2, Fig. \ref{Fig:VTTallmaps}b) are displayed in the upper panel of Fig. \ref{Fig:Hevlos}. The solid and dotted crosses in the upper panel mark the positions of the spectral profiles shown in the lower panel. The \ion{He}{i} triplet appears broadened, clearly redshifted, and strongly in emission. The gray line shows the quiet-Sun profile, while the red line shows the fit to the flare profile retrieved from the HAZEL inversions. The fits are accurate and the obtained velocities within the flaring area are dominated by redshifts. 

Interestingly, surrounding the redshifted flare area, at the front of the flare ribbons, the velocities are blueshifted. After inspection of the spectral profiles (discussed in Sect. \ref{Sect:DiscLeadingfFront}), we conclude that these blueshifts are real, as the profiles exhibit a blueshifted absorption component that co-exists with another, weaker, emission component. 
Two-component spectral inversions would be necessary to fit this redshifted emission profile. The co-existence of blue and redshifted \ion{He}{i} profiles is interesting in itself and shows that this phenomenon is common in pixels affected by the leading edge of the flare. We will come back to this in the discussion.

\subsection{Ambiguity in the velocities during the peak} \label{Sect:ambiguity}
The ambiguity seen in the \ion{Ca}{ii} 8542\,\AA\ velocities of Fig.~\ref{Fig:caiivelmaps} can be solved by comparing them to the \ion{He}{i} 10830\,\AA\ velocities in Fig.~\ref{Fig:Hevlos}. The main bright flaring area, which is dominated by emission profiles, is uniformly redshifted for the \ion{He}{i} velocities, apart from the narrow region at the ribbon front. Since \ion{He}{i} forms higher in the atmosphere, a solution in which \ion{Ca}{ii} is also redshifted seems more plausible. In addition, a visual inspection of the example profiles in Fig. \ref{Fig:Caprofiles} reveals shifts of the line core toward the red.
Therefore, configuration A from Table \ref{Tab:nodes} of the \ion{Ca}{ii} inversions is strongly favored. The results therefore must be interpreted with caution, ideally using a second spectral line in the chromosphere if available, like in the present study. 
Another interpretation, although inconsistent with our discussion, is based on the inversion code's results from configuration B, which identify the lower-lying left wing of the \ion{Ca}{ii} 8542\,\AA\ line (see Fig.~\ref{Fig:Caprofiles}) as a blueshifted absorption profile.
For the remainder of this study, we present only the \ion{Ca}{ii} LOS velocities obtained from the simpler inversions using configuration A of Table \ref{Tab:nodes}.

\begin{figure*}[!t]
 \sidecaption
 \includegraphics[width=0.7\hsize]{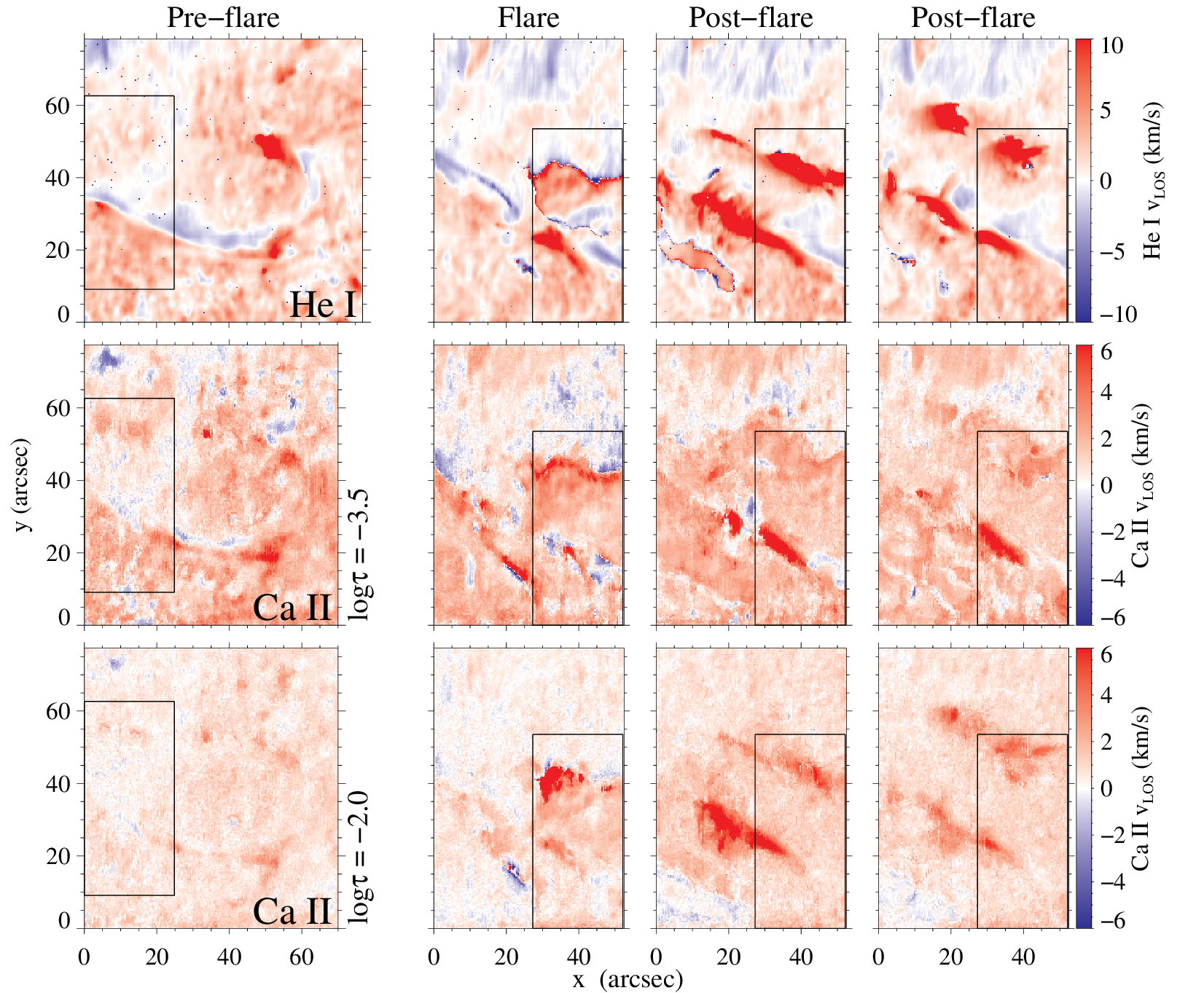}
  \caption{LOS velocities at different times of the flare inferred from the inversions. From top to bottom: \ion{He}{i} 10830\,\AA, \ion{Ca}{ii} 8542\,\AA\ at $\log\tau = -3.5$, and $\log\tau = -2.0$. The first column of velocity maps corresponds to the pre-flare phase (07:49--08:29 UT). The second column shows the peak of the flare (08:35--09:05 UT), whereas the third and fourth columns depict the post-flare phases corresponding to the time ranges (09:07--09:36 UT) and (09:38--10:08 UT), respectively. Note that the top panels corresponding to the \ion{He}{i} LOS velocities are clipped between $\pm10$\,\kms, whereas the middle and lower panels of \ion{Ca}{ii} LOS velocities are saturated between $\pm6$\,\kms. The black box shows the common FOV among all raster scans. } 
  \label{Fig:velocitymaps}
\end{figure*}

\subsection{Temporal evolution}
The LOS velocities for the different phases of the M-class flare are exhibited in Fig. \ref{Fig:velocitymaps}. The upper panels show the velocities retrieved from the \ion{He}{i} inversions, whereas the middle and lower panels show the velocities inferred from the \ion{Ca}{ii} inversions at $\log\tau = -3.5$ and $-2.0$. 
For a better comparison in time, we focus on the black rectangles, which show the common FOV across the four maps. The pre-flare map shows blueshifts associated to the central filament in the upper \ion{He}{i} panel. These upflows are also seen, but much weaker, in the \ion{Ca}{ii} at the $\log\tau = -3.5$ layer. Surrounding the filament we mainly see redshifts. 

The second column in Fig. \ref{Fig:velocitymaps} illustrates the LOS velocities during the M3.2 flare. Within the black box, we see downflows in the central part, which belong to the main flare ribbon, among all layers. These redshifts are linked to the largest emission profiles of \ion{He}{i} and \ion{Ca}{ii}. Interestingly, a narrow area at the front of the flare ribbon, also called the leading edge of the flare, shows blueshifts. This is nicely seen in the \ion{He}{i} velocity map, and is less pronounced but consistent, in some parts of the \ion{Ca}{ii} velocity maps. The \ion{He}{i} velocities at this blueshifted thin layer are between $-4$ and $-10$\,\kms. The \ion{Ca}{ii} velocities at $\log\tau = -3.5$ show a well-enhanced patch of blueshifts as well, at the upper-right leading flaring front at about ($x,y = 47\arcsec,44\arcsec$). The velocities are between $-2$ and $-5$\,\kms, and much lower of about $-1$\,\kms\ at $\log\tau = -2.0$. The small southern ribbon ($x,y = 25\arcsec,15\arcsec$) exhibits mostly blueshifts in \ion{He}{i} and \ion{Ca}{ii} at $\log\tau = -2.0$, whereas \ion{Ca}{ii} at $\log\tau = -3.5$ shows blueshifts only at the southern edge of that ribbon. 

The filament partially remains blueshifted, with larger parts redshifted in \ion{Ca}{ii}. No eruption of the filament takes place at any time. The redshifted flare front of the large central ribbon, next to the leading blueshifted edge, shows LOS velocities in \ion{He}{i} in the range of $10-13$\,\kms. LOS velocities between $5-9$\,\kms are found in the upper ribbon at the flare front for \ion{Ca}{ii} at $\log\tau = -3.5$, whereas at $\log\tau = -2.0$ the velocities are even larger, between $10-25$\,\kms. The central part of the ribbon (around $x,y \sim 35\arcsec,38\arcsec$) exhibits slower velocities, in the range of 2--7\,kms\ for \ion{He}{i} and \ion{Ca}{ii} at $\log\tau = -2.0$, and 2--5\,kms\ at $\log\tau = -3.5$.

The last two post-flare maps display extensive downflows on both sides of the filament, featuring redshifted profiles. These downflows occur at the footpoints of a loop system that crosses the filament and the PIL, as shown in a magnetogram by \citet{kuckein15b}, with its roots in different polarities.
The loop system is clearly seen in the \ion{He}{i} maps in Fig. \ref{Fig:VTTallmaps} (panels c--d). The redshifts are most prominent in \ion{He}{i} on both sides of the filament. In the \ion{Ca}{ii} velocities, the southern side exhibits much stronger redshifts, while the upper part shows only minor shifts at $\log\tau = -2.0$. Although the \ion{He}{i} maps suggest that plasma flows consistently downward on both sides of the filament, the \ion{Ca}{ii} maps indicate a clear preference for the southern side. The \ion{He}{i} velocities in the southern footpoint are between $20-30$\,\kms, whereas the upper footpoint has areas of up to 39\,\kms, in the first post-flare map. In the second post-flare map, the area of strong redshifts decreases, though some pixels of about 30\,\kms\ still exist in both footpoints. The \ion{Ca}{ii} post-flare maps exhibit downflows in the southern footpoints. 
At $\log\tau = -3.5$, only a small patch is visible, which appeared shortly after the flare at approximately ($x,y) \sim (22\arcsec,27\arcsec$), with LOS velocities in the range of $6-10$\,\kms. In the second post-flare map the patch vanished. The lowest atmospheric layer, given by $\log\tau = -2.0$ of the \ion{Ca}{ii} line, shows similar flows in the southern footpoint, in the range of $7-9$\,\kms\ with peaks slightly exceeding 10\,\kms\ for the first post-flare map. In the second post-flare map, the velocities have weakened considerably, and both footpoints exhibit redshifted LOS velocities of about $3-5$\,\kms\ with peaks of $\sim 6$\,\kms\ in the northern footpoint.

\section{Discussion}

\subsection{Height sensitivity of the \ion{Ca}{ii} line}
The computed response functions for \ion{Ca}{ii} 8542\,\AA\ in the flare ribbon indicate a shift in spectral line sensitivity to Doppler shifts toward lower heights (Fig. \ref{Fig:RF}). This affects the result interpretation, as the \ion{Ca}{ii} line core is typically expected to form higher in the atmosphere.  

\citet{kuridze18} computed RFs for \ion{Ca}{ii} in response to temperature and LOS magnetic-field changes in an M-class flare. Consistent with our M3.2 flare results, they found an optical depth shift to lower heights, closer to the photosphere during the flare. This effect is also observed, but less prominent, in smaller flares. \citet{yadav21} demonstrated in their Fig. 11 that, in a C2 flare, the \ion{Ca}{ii} line wings become sensitive down to $\log \tau \sim -2$, while the line core shifts down by approximately $\Delta\log \tau \sim 0.5$ in optical depth.  

All the abovementioned studies, including ours, consistently show a decrease in the height of \ion{Ca}{ii} line sensitivity. \citet{kerr16} found the same behavior in simulations, attributing it to high temperatures that reduce the upper-level population of \ion{Ca}{ii} 8542\,\AA, shifting the formation height downward. We emphasize that \ion{Ca}{ii} is ionized to \ion{Ca}{iii}, making it undetectable in the \ion{Ca}{ii} 8542\,\AA\ observations. This implies that the analysis should be restricted to higher optical depths (Fig. \ref{Fig:RF}), that is lower heights, than is typically done in active-region studies with this spectral line. 
It is also worth noting that a downward shift in the height of line sensitivity has also been observed in other spectral lines during flares. In particular, the \ion{He}{i} 10830\,\AA\ and \ion{He}{i} D$_3$ lines can exhibit a similar behavior \citep[][]{anan18, libbrecht19}.

\begin{figure*}[!t]
 \includegraphics[width=\hsize]{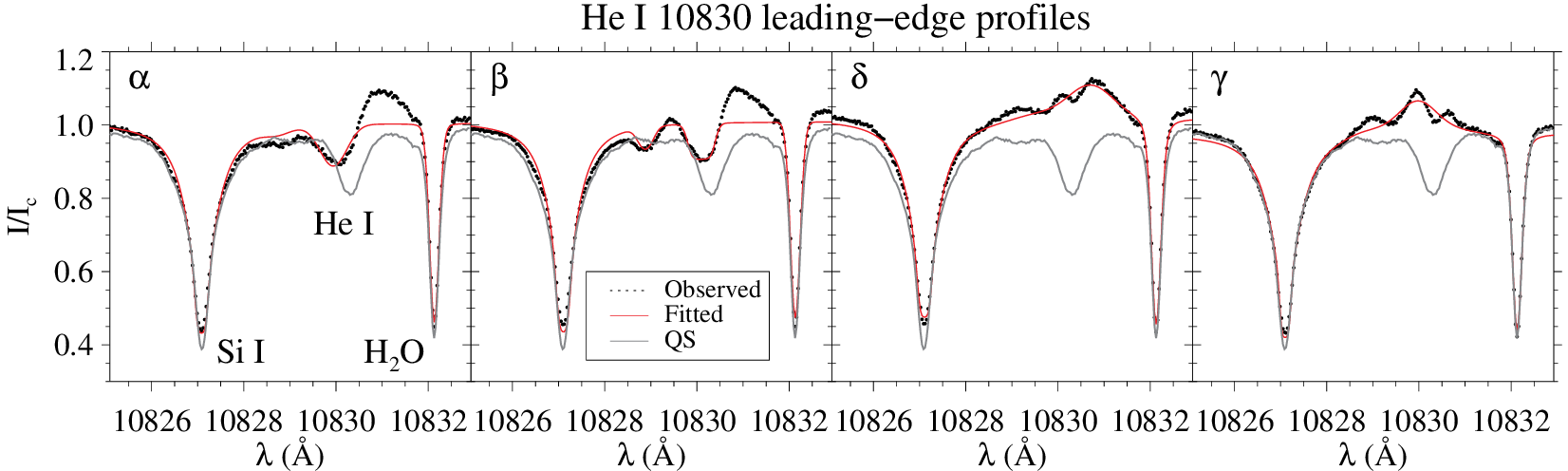}
  \caption{Examples of \ion{He}{i} 10830\,\AA\ intensity profiles found at the leading front of the flare ribbon. Dots represent the observed profiles and the red solid line is the fit from the HAZEL inversions. The quiet-Sun profile, which is at rest, is shown in gray to identify better the Doppler shifts. The position of the four profiles ($\alpha$, $\beta$, $\delta$, and $\gamma$) appear in Fig. \ref{Fig:Hevlos}. The inferred velocities are ($\alpha - \gamma$): $-10.0$, $-5.8$, $10.9$, and $-9.1$\,\kms, respectively. In panel $\alpha$, the absorption seen in \ion{He}{i} is the blueshifted component, whereas the \ion{He}{i} emission profile is the redshifted component. } 
  \label{Fig:profsbluefront}
\end{figure*}

\subsection{Leading front of the flare} \label{Sect:DiscLeadingfFront}
The present M3.2 flare observations reveal a leading front of the ribbon dominated by chromospheric blueshifts. This is most prominent in the \ion{He}{i} 10830\,\AA\ velocity map during the flare (Fig. \ref{Fig:velocitymaps}) and remains visible, though less pronounced, in the first (lower-left area) and second post-flare maps. The \ion{Ca}{ii} 8542\,\AA\ velocity maps also exhibit this behavior, but the response is weaker, with less prominent blueshifts. Next to the leading edge of the flare, the bright part with clearly redshifted emission profiles follows. The blueshifted flare front in the \ion{He}{i} observations stems from the presence of two distinct velocity components within each pixel -- one blueshifted and one redshifted -- with our inversions predominantly favoring the blueshifted component. Figure \ref{Fig:profsbluefront} exhibits four example profiles of \ion{He}{i} 10830\,\AA\ from the leading front, surrounding the bright flare area. Although pixel $\delta$ is mostly redshifted, there is also a shallow blueshifted \ion{He}{i} profile, which the inversion code did not fit. 
During the inversion process of the spectral profiles, the inversion code mostly favored the blueshifted profile, producing a homogeneous blueshifted area in space surrounding the bright redshifted emission profiles (Figs. \ref{Fig:velocitymaps} or \ref{Fig:Hevlos}). The coexistence of two components may indicate that two mechanisms of line formation are at work, or even the transition from one mechanism to the other due to the flare. On the one hand, the triplet can be populated by increased EUV photoionization due to the flare and subsequent recombinations \citep{mauas05,centeno08}. On the other hand, non-thermal collisional ionization and recombination can generate \ion{He}{i} \citep{ding05}. Both mechanisms seem reasonable to be enhanced during flares when there is evidence for non-thermal electrons in the chromosphere. \citet{kerr21} found that the latter mechanism was necessary to produce flare-induced dimming of \ion{He}{i} 10830\,\AA. 
We likely see this dimming only at the north front of the flare ribbon in Fig. \ref{Fig:VTTallmaps}b), which is typically seen as a darkening in \ion{He}{i} line-core images. \citet{xu16} also reported dimmings at only one side of the flare ribbon. 
The reason for not seeing it clearly in our observations might be the lack of spatial resolution needed to identify these dimmings of about 340\,km ($\sim0\farcs47$), as reported by \citet{xu16}. Our theoretical spatial resolution is $\sim0\farcs70$. Nevertheless, dimming is produced by having \ion{He}{i} absorption profiles, which we found shifted to the blue in the flare front (e.g., pixels $\alpha$ and $\beta$ in Fig. \ref{Fig:profsbluefront}). The lifetime of the dimming is about 91\,s \citep{xu16}. Our data does not have sufficient time resolution to measure the duration of our dimming -- the raster scan took about 30\,min. 
However, the presence of blueshifted absorption profiles alongside emission profiles within the same pixel suggests a rapid transition to redshifted emission profiles in the bright area as the flare evolves.

A blueshifted flare front has been reported in other chromospheric lines \citep{panos18}. \citet{polito23} reported \ion{Mg}{ii} k3 blueshifts at the leading edge of an X-class flare (their Fig. 4). Moreover, the authors found that the optically thin transition region line \ion{O}{iv} was also blueshifted in some locations of the flare front. \cite{tei18} showed \ion{Mg}{ii} blueshifted lines of about $-10$\,\kms, which evolved into redshifts at the flare front of a C5.4 flare. The authors propose a cool-upflow scenario (their Fig. 12) to explain this phenomenon: high-energy electrons reach deeper layers, heating the plasma to temperatures of the order of $10^6$\,K. As a result, the cooler chromospheric plasma above the heated region is pushed upward, appearing briefly as a blueshift in chromospheric spectral lines until the intense radiation from the condensation region dominates. Although our data lack the temporal resolution needed to follow the evolution of the profiles, we find blue and redshifted profiles within the same pixel, indicating that a very short time may pass in the transition from blue to redshifted profiles. The lifetime of the blueshifted \ion{Mg}{ii} profiles presented by \cite{tei18} is, on average, 29\,s. 

Numerical experiments on the ribbon front in flares support the scenario of high-energy electrons penetrating deeper layers, primarily perturbing the mid-chromosphere \citep{kerr24}. Evaporation then takes place producing upflowing plasma -- the blueshifted leading edge. At some point, the upper chromosphere gets strongly affected by the flare and plasma flows become redshifted, producing the emission areas seen in \ion{He}{i} and \ion{Ca}{ii} (Fig.~\ref{Fig:VTTallmaps}). This explains why our \ion{Ca}{ii} velocities exhibit much weaker blueshifts at the ribbon front (Fig.~\ref{Fig:velocitymaps}), because the line is formed deeper in the atmosphere than \ion{He}{i}. 

Our \ion{He}{i} 10830\,\AA\ data reveal a blueshifted leading front, a feature whose interpretation remains open in the context of our event. It could be attributed either to rapid chromospheric heating \citep{tei18} or to gentle evaporation \citep{kerr24}. A detailed thermal analysis could help clarify this question, but it is deferred to a future study, as it lies beyond the scope of the present work.

\subsection{\ion{Ca}{ii} 8542\,\AA\ line shape and shifts}
It is worth discussing the line shape of the \ion{Ca}{ii} 8542\,\AA\ line during the flare. The current data show the line at a much finer spectral sampling (25\,m\AA) compared to flare studies based on imaging-spectroscopy instruments. Figure \ref{Fig:Caprofiles} shows an example of four intensity profiles selected close to the front of the flare ribbon. They show clear emission profiles without a central reversal of the line core -- as shown in simulations \citep{kerr16} -- and with peaks of about $I/I_\mathrm{c} \sim 1.7$. In contrast, \citet{kuridze18} showed in their Fig. 6 averaged \ion{Ca}{ii} 8542\,\AA\ flare profiles with a shallow dip in the line-core center, in an M1.9 flare. Moreover, \citet{kuridze17} reported a weaker emission profile, without central reversal, in a C8.4 flare. \citet{ferrente24} showcased two \ion{Ca}{ii} profiles inside the flare ribbon of an X-class flare. One profile reaches a peak of about $I/I_\mathrm{c} \sim 2.0$, without a central reversal. \citet{dacosta16} also presented \ion{Ca}{ii} profiles in their Fig. 3 during an X-class flare, showing no central reversal in the flare ribbon.
Our profiles seem compatible but show more details in the core and wings due to the better spectral resolution. All profiles from the flare front shown in Fig.~\ref{Fig:Caprofiles} have an asymmetric shape, with a predominant broader red wing indicating velocity gradients and downward plasma motions. Interestingly, the peak intensity is not at the center of the core of the line. \citet{kerr16} found in their simulations of the \ion{Ca}{ii} 8542\,\AA\ line during flare heating a tiny shifted component in the core. They interpreted that as optically thin emission with a higher temperature than the rest of the core. Nevertheless, because the line is optically thick, its peak intensity does not necessarily correspond to the core of the line. In Fig. \ref{Fig:Caprofiles}, the small optically thin emission within the core is always found towards the blue. However, the inversion code can hardly account for these tiny shifts inside the core. Therefore, the impact on the inversions seems rather negligible. We note that the flare profiles from an M1.1 flare shown by \citet{kuridze15} in their Fig. 3 might also indicate a slightly blueshifted peak inside the core of the line. Nevertheless, their spectral sampling is much too broad (100\,m\AA) to confirm this peak reliably. 

\subsection{Plasma-flow and flare-loops evolution during the flare}
Figure \ref{Fig:velocitymaps} shows the evolution of the LOS velocities at different stages of the flare and at several heights. It is worth keeping Fig. \ref{Fig:VTTallmaps} in mind to compare the velocities to their intensity raster scans. The impulsive phase of the flare (second column in Fig.~\ref{Fig:velocitymaps}) leaves an important imprint in the chromosphere. Not only do the \ion{He}{i} 10830\,\AA\ and \ion{Ca}{ii}  8542\,\AA\ profiles appear in emission (e.g., Figs. \ref{Fig:Caprofiles}, \ref{Fig:Hevlos}, and \ref{Fig:profsbluefront}), but also most of these profiles are redshifted. Hence, we see plasma flows toward the solar surface in the heated part and across all layers. We note a faint blueshifted region ($x,y \sim 40\arcsec, 30\arcsec$) in the \ion{He}{i} velocity panels in Fig. \ref{Fig:velocitymaps}, located within the central flare ribbon and inside the black square. This flow remains spatially and temporally persistent in the subsequent post-flare maps. Therefore, it seems not to be part of the flare ribbon itself, which moves away from the PIL, but could show long-lasting slow evaporation. 

Although redshifts dominate the flare ribbon in our data, a recent study of an X-class flare by \citet{ferrente24} reports predominantly blueshifts all over the ribbon at $\log \tau = -4$ with velocities about $-20$\,\kms. 
We raise the question of whether the strong upflows at $\log \tau = -4$ could result from ambiguous solutions in the inversion code, similar to those discussed in Sect.~\ref{Sect:ambiguity}. 
These upflows represent indeed a well-defined solution to the fitting problem, particularly given the significant sensitivity of the RF at that layer in their study. In our case, an examination of the spectral profiles revealed that not only do the flare-emission profiles exhibit a red asymmetry, but the inner core of the line is also shifted toward the red. We speculate that with fewer spectral points, as in their observations, the inversion code might be more prone to finding alternative solutions for the same profiles. Notably, identifying different solutions requires running the computationally intensive inversion process multiple times. This question has been raised before; since the 1980s, researchers have reported observations of areas within ribbons with strong red-asymmetries, suggesting that these features are more likely due to the downward motion produced by the impulsive heating than attenuation in the blue wing due to rising matter over the flare \citep{Ichimoto1984}. This highlights the importance of multiple chromospheric lines to resolve such ambiguities, like in the present study, and new diagnostic tools that can capture the multimodality of the solution \citep{2022A&A...659A.165D}.

In addition, as highlighted in Sect. \ref{Sect:DiscLeadingfFront}, a blueshifted leading edge of the flare ribbon is seen, more prominently in the \ion{He}{i} velocities. We assume that chromospheric evaporation, as described by \citet{tei18} and \citet{kerr24}, is the reason for these short-lived blueshifts. Blueshifts are predominantly observed at the northern edge of the flare ribbon, consistent with previous studies that have reported enhanced absorption in only one flare ribbon.

Following the ionization of \ion{He}{i} 10830\,\AA\ due to the flare, recombination sets in, producing strong absorption in the flare-affected region. The first post-flare velocity map (column 3 in Fig. \ref{Fig:velocitymaps}) of \ion{He}{i} 10830\,\AA\ reveals strong downflows on both sides of the filament, and hence the PIL. The intensity raster scan reveals dense \ion{He}{i} absorption following loop-like arches crossing the filament, connecting opposite polarities. 
Interestingly, the loop tops do not exhibit significant velocities, suggesting that these loops are predominantly at rest and are not expanding at this stage of the flare (about 30\,min after the impulsive phase). The plasma flows at the footpoints of the loops are strongly enhanced, suggesting plasma flowing downward following these loops. Notably, \ion{Ca}{ii} 8542\,\AA\ does not display loop-like structures in Fig. \ref{Fig:VTTallmaps}, likely because most \ion{Ca}{ii} is ionized at the higher loop-top regions. The strongest downflows appear at the footpoints of the loops in both polarities for \ion{He}{i}, whereas \ion{Ca}{ii} exhibits strong redshifts only at one side of the PIL at $\log \tau = -3.5$. These downflows at the footpoints persist, as they are also observed in the one-hour-later post-flare velocity map for both \ion{He}{i} and \ion{Ca}{ii} at $\log \tau = -2.0$.

A persistent strong redshifted area, with about 10\,\kms, draws our attention at the lower left part within the black box in Fig. \ref{Fig:velocitymaps}, at all stages of the flare, and across all shown heights ($x,y \sim 30\arcsec, 23\arcsec$). This area is outside of the flare ribbon. The redshifts start at the impulsive phase of the flare -- but less pronounced in \ion{Ca}{ii} compared to \ion{He}{i} -- and are present for at least one hour. The raster scans in Fig. \ref{Fig:VTTallmaps} indicate that the redshifts are close to the filament. We therefore speculate that these redshifts are related to the filament, which remains stable during the flare. The literature also contains examples of active region filaments that persist after a flare occurs over them \citep[e.g.,][]{DiazBaso_2019F}, emphasizing the importance of magnetic connectivity during the evolution of these events.

\section{Summary and conclusions}
We presented the chromospheric LOS velocities associated with an M3.2 flare, as inferred from the combined analysis of two different spectral regions: the \ion{He}{i} 10830\,\AA\ triplet and the \ion{Ca}{ii} 8542\,\AA\ line. The ground-based raster observations at VTT covered the pre-flare, flare, and post-flare phases. 

Our main findings are:
\begin{enumerate}
  \item The response function analysis reveals that the \ion{Ca}{ii} 8542\,\AA\ line forms at lower heights during flares, with peak sensitivity shifting from $\log\tau \approx -5.2$ to $-3.5$ in our M3.2 flare. These results are consistent with the limited studies that have investigated \ion{Ca}{ii} RFs in other flare classes.

  \item Using multiple chromospheric spectral lines is crucial for resolving inversion ambiguities in the LOS velocities, as demonstrated by our comparison between \ion{He}{i} and \ion{Ca}{ii} velocities (Sect. \ref{Sect:ambiguity}).

  \item The flare exhibits blueshifts at the narrow leading edge of the ribbons -- best seen at  \ion{He}{i} 10830\,\AA\ with LOS velocities of up to $-10$\,\kms\ -- and redshifts in the main bright flaring areas across all spectral lines with velocities between $2-13$\,\kms\ in \ion{He}{i}. The blueshifted leading-edge profiles mostly reveal a coexistence of blueshifted absorption and redshifted emission components in \ion{He}{i} 10830\,\AA\ (see Fig. \ref{Fig:profsbluefront}). This supports a cool-upflow scenario with rapid transition from upflows to downflows, likely occurring within seconds to tens of seconds. 
  
  \item Dimming in \ion{He}{i} is likely seen on only one side of the flare ribbon, similar to the few other \ion{He}{i} observations during flares. 
  
  \item Post-flare observations show loop-like structures in \ion{He}{i} connecting opposite polarities with strong downflows at footpoints (up to 39\,\kms), while \ion{Ca}{ii} exhibits asymmetric downflows, much stronger in the southern footpoint. The \ion{Ca}{ii} downflows are up to 10\,\kms\ for both, $\log\tau = -3.5$ and $\log\tau = -2.0$. No loop-like structures are seen in \ion{Ca}{ii}. 

  \item The filament remained stable throughout the flare, showing no signs of eruption.
\end{enumerate}

This study demonstrates the importance of combining spectroscopic observations from different chromospheric lines to create a comprehensive picture of flare dynamics. This is particularly important for understanding the complex velocity fields during different phases of solar flares. More such high-resolution multi-wavelength observations are crucial to further understanding the chromospheric dynamics during flares. The advent of new-generation solar telescopes such as DKIST \citep{DKIST}, EST \citep{EST}, and WeHot \citep{WeHot} will be crucial in this regard.

\begin{acknowledgements}
The Vacuum Tower Telescope is operated by the Institute for Solar Physics in Freiburg, Germany, at the Spanish Observatorio del Teide, Tenerife, Canary Islands. This project has received funding from the European Union's Horizon 2020 research and innovation programme under the Marie Sk\l{}odowska-Curie grant agreement No 895955. 
We acknowledge support from the Agencia Estatal de Investigación del Ministerio de Ciencia, Innovación y Universidades (MCIU/AEI) under grant ``Polarimetric Inference of Magnetic Fields'', the European Regional Development Fund (ERDF) with reference PID2022-136563NB-I00/10.13039/501100011033.
CK acknowledges grant RYC2022-037660-I funded by MCIN/AEI/10.13039/501100011033 and by "ESF Investing in your future".
MC acknowledges financial support from Ministerio de Ciencia e Innovacion and the European Regional Development Fund through grant PID2021-127487NB-I00.
This research is supported by the Research Council of Norway, project number 325491, 
and through its Centres of Excellence scheme, project number 262622. 
TF acknowledges grants PID2021-127487NB-I00, CNS2023-145233 and RYC2020-030307-I funded by MCIN/AEI/10.13039/501100011033.
CQN acknowledges Project ICTS2022-007828, funded by MICIN and the European Union NextGeneration EU/RTRP.
LK acknowledges support from an SNSF PRIMA grant and a SERI-funded ERC CoG grant.
LF acknowledges support from UK Research and Innovation’s Science and Technology Facilities Council grant ST/X000990/1.
SM acknowledges support from STFC grants ST/W001004/1 and ST/V003658/1.
Drs. H. Socas-Navarro, B. Ruiz Cobo, and E. S. Carlin are greatly acknowledged for fruitful discussions.
 
\end{acknowledgements}


\bibliographystyle{aa}
\bibliography{aa-jour,biblio}

\end{document}